\documentclass[vecphys]{svmult}

\usepackage{makeidx}         
\usepackage{graphicx}        
\usepackage{multicol}        
\usepackage[bottom]{footmisc}
\makeindex             

\newcommand{\be}{\begin{equation}}    
\newcommand{\ee}{\end{equation}}
\newcommand{\ba}{\begin{eqnarray}}
\newcommand{\ea}{\end{eqnarray}}

\begin{document}

\title*{Hamiltonian Normal Forms and Galactic Potentials}
\author{Giuseppe Pucacco\inst{1,2}}
\institute{Dipartimento di Fisica -- Universit\`a di Roma ``Tor Vergata", Via della Ricerca Scientifica, 1 -- 00133 Rome, Italy
\texttt{pucacco@roma2.infn.it}
\and INFN - Sezione di Roma Tor Vergata}
\maketitle

\section{Introduction}
\label{sec:1}

The study of self-gravitating stellar systems has provided important hints to develop tools of analytical mechanics. We may cite the ideas of Jeans \cite{jeans} about the relevance of conserved quantities in describing the phase-space structure of large N-body systems and his introduction of the concept of {\it isolating integral}. Later important contributions are those of Contopoulos \cite{c1}, who applied a direct approach to compute approximate forms of the isolating integrals of motion, of H\'enon \& Heiles \cite{hh} with a paradigmatic example of non-integrable system derived from a simple galactic model and of Hori \cite{hori}, who introduced the theory of Lie transforms in the field of canonical perturbation theory. These and other cues contributed to the body of methods and techniques that we use today to study regular and chaotic rdynamics of non-integrable systems. 

The direct approach applied by Contopoulos aims at solving the equation for the conserved quantity along the classical procedure developed early in the last century \cite{wh}. The method of the Lie transform \cite{hori}, subsequently improved by several authors \cite{deprit,kamel,kummer,gg,giorgilli}, has some technical advantages and has gradually become a standard method in the perturbation theory of Hamiltonian dynamical systems \cite{bp}. However, its application in galactic dynamics, with a few remarkable exceptions \cite{gs,ya}, has not been as systematic and productive as it could be.

Hamiltonian normal forms constructed in this way \cite{BBP1, BBP2, BBP3} are a powerful tool to investigate the orbit structure of galactic potentials and to gather several qualitative informations concerning the near integrable dynamics below the stochasticity threshold (if any) of the system. Results obtained in the same class of systems by the averaging method \cite{zm} are easily overtaken. As a matter of fact, with a normal form truncated to an order sufficient to incorporate the main resonance, one can also make reliable {\it quantitative} predictions. In the present contribution we review how to exploit detuned resonant normal forms to extract information on several aspects of the dynamics in systems with self-similar elliptical equipotentials. In particular, using energy and ellipticity as parameters, we compute the instability thresholds of axial orbits, bifurcation values of low-order boxlets and phase-space fractions pertaining to the families around them. We also show how to infer something about the singular limit of the potential. 

A remarkable side-effect of expressing the stability--instability threshold as a series expansion, is that its predictive ability goes well beyond the radius of convergence of the perturbing expansion. Exploiting asymptotic properties of the series constructed via the normal form \cite{CEG, ECG}, we may try to estimate an optimal truncation order.

\section{The Hamiltonian normal form}
\label{sec:2}

The subject of our investigation is the class of 2-dof natural systems
\be\label{Horig}
   H(\vec{p},\vec{r})= \frac{1}{2}(p_x^2+p_y^2 / q) + V(s(x,y)).
\ee
$V$ is a uniformly increasing function of the variable
\be\label{ellipse}
s = x^2 + y^2 / q,\ee
with an absolute regular minimum ($V(0,0)=V'(0,0)=0$), so that the energy $E$ may take any non-negative value. Two simple examples are 
\ba
V_{L}&=&\frac12 \log(1 + s),\label{v1}\\
V_{C}&=&\sqrt{1 + s}-1.\label{v2}\ea
 The parameter $q$ gives the ``ellipticity'' of the equipotentials and ranges in the interval
 \be\label{rangeq}
 0.6 < q < 1. \ee
 Lower values of $q$ can in principle be considered but correspond to an unphysical density distribution if $V$ is a gravitational potential. Values greater than unity are included in the treatment by reversing the role of the coordinate axes. With respect to the standard `physical' notation, the scaling transformation
\be
p_y \longrightarrow {\sqrt{q}} \, p_{y}, \quad  
y \longrightarrow y/\sqrt{q}\ee
is implicit in the Hamiltonian written in the form (\ref{Horig}).

\subsection{Series expansions}
\label{subsec2:1}

To investigate the dynamics of system (\ref{Horig}), we look for a new Hamiltonian given by the series expansion in the new canonical variables $\vec{P},\vec{R}$, 
\begin{equation}\label{HK}
     K({\vec{P},\vec{R}})=\sum_{n=0}^{\infty}K_n ({\vec{P},\vec{R}}),
  \end{equation}
with the prescription that 
\be\label{NFD}
\{H_0,K\}=0.
\ee
In these and subsequent formulas we adopt the convention of labeling the first term in the expansion with the index zero: in general, the `zero order' terms are quadratic homogeneous polynomials and terms of {\it order} {\it n} are polynomials of degree $n+2$. The zero order (unperturbed) Hamiltonian,  
\be\label{Hzero}
H_{0} ({\vec{P},\vec{R}})\equiv K_{0} = \frac12 (P_X^2 + X^2) + \frac{1}{2q} (P_Y^2  +Y^2),
\ee
with unperturbed frequencies $\omega_1=1$ and $\omega_2 = 1/q$, is expressed in terms of the new variables found at each step of the normalizing transformation. 
It is customary to refer to the series constructed in this way as a ``Birkhoff'' normal form \cite{bir}. The presence of terms with small denominators in the expansion, forbids in general its convergence. It is therefore more effective to work since the start with a {\it resonant normal form} \cite{SV}, which is still non-convergent, but has the advantage of avoiding the small divisors associated to a particular resonance. To catch the main features of the orbital structure, we therefore approximate the frequencies with a rational number plus a small ``detuning'' 
\be\label{DET}
\frac{\omega_1}{\omega_2} = q = \frac{m_1}{m_2} + \delta. \ee
We speak of a {\it detuned} ($m_1/m_2$) {\it resonance}, with $m_1+m_2$ the {\it order} of the resonance. In order to implement the normalization algorithm, also the original Hamiltonian (\ref{Horig}) has to be expressed as a series expansion around the equilibrium: performing
the rescaling
\be\label{newE}{\cal H} := \frac{m_2 H}{\omega_2} = m_2 q H,\ee
we redefine the Hamiltonian as the series
\be\label{detH}
{\cal H} = \sum_{k=0}^{\infty} {\cal H}_{k} = \frac12 [m_1 (p_x^2+x^2) + m_2 (p_y^2+y^2)]+{\scriptstyle{\frac12}} m_2\delta (p_x^2+x^2) + \sum_{k=1}^{\infty} b_{k} (q) s^{k+1} ,\ee 
with expansion coefficients $b_{k}$ depending only on the ellipticity in view of the restriction imposed by the choice of the potentials. The procedure is now that of an ordinary resonant ``Birkhoff--Gustavson'' normalization \cite{gu, moser} with two variants: the coordinate transformations are performed through the Lie transform and the detuning quadratic term is treated as a term of higher order and put into the perturbation. This is analogous to the strategy of the `nearly resonant construction' of Contopoulos and 
Moutsoulas \cite{CM} in the context of the direct approach and is implemented in the program by Giorgilli \cite{gprog}. 

\subsection{Lie transform normalization}
\label{subsec2:2}

Considering a generating function $g$, the new coordinates $\vec{P},\vec{R}$ result from the canonical transformation
  \be\label{TNFD}
  (\vec{P},\vec{R}) = M_{g} (\vec{p},\vec{r}).\ee
The {\it Lie transform operator} $M_{g}$ is defined by \cite{bp}
\begin{equation}\label{eqn:OperD-F}
    M_{g} \equiv \sum_{k=0}^{\infty} M_k
\end{equation}
where
\be M_0 = 1, \quad M_k = \sum_{j=1}^k \frac{j}{k} L_{g_j} M_{k-j}.\ee
The functions $g_j$ are the terms in the expansion of the generating function ($g_{0}=1$)
and the linear differential operator $L_{g}$ is defined through the Poisson bracket, $L_{g}(\cdot)=\{g,\cdot\}$.

The terms in the hew Hamiltonian are determined through the recursive set of linear partial differential equations \cite{bp}
\be\label{EHK} 
K_n= {\cal H}_n +\sum_{j=0}^{n-1}M_{n-j}{\cal H}_j ,
\;\; n=1,2, \dots
\ee 
`Solving' the equation at the $n$-th step consists of a twofold task: to find $K_{n}$ {\it and} $g_n$. We observe that, in view of the reflection symmetries of the Hamiltonian (\ref{Horig}), the chain (\ref{EHK}) is composed only of members with even index and so the normal form itself is composed of even-index terms only. The unperturbed part of the Hamiltonian, ${\cal H}_0$, determines the specific form of the transformation. In fact, the new Hamiltonian $K$ is said to be {\it in normal form} if, analogously to (\ref{NFD}),
\be
\{{\cal H}_0,K\}=0,
\ee
 is satisfied. The function 
\be\label{integral}
{\cal I} = K - {\cal H}_{0} \ee
can be used as a second integral of motion. For practical applications (for example to compare results with numerical computations) it is useful to express approximating functions in the original physical coordinates. Inverting the coordinate transformation, the new integral of motion can be expressed in terms of the original variables. Denoting it as the power series
\be\label{PHI}
 I = \sum_{n=0}^{\infty}I_n,\ee
its terms can be recovered by means of the set of equations
\be
\sum_{j=1}^{n}M_{n-j}\big[{\cal H}_j-I_j\big] -K_n = 0, \qquad n\geq 1 ,
\label{eqn:In}\ee
that is obtained from (\ref{EHK}) and (\ref{integral}) by exploiting the nice properties of the Lie transform with respect to inversions \cite{bp}.

\subsection{Effective order of the detuning}
\label{subsec2:3}

We have to discuss how to treat the detuning term: it is considered as a higher order term and the most natural choice is to put it into ${\cal H}_2$. However, there is no strict rule for this and one may ask which is the most `useful' choice, always considering that applications are based on series expansions with coefficients depending on $q$. We remark that, different choices of the {\it effective order}, say $d$, of the detuning, lead to different terms of higher order in the normal form. We also observe that, whatever the choice made, the algorithm devised to treat, step by step, the system (\ref{EHK}) must be suitably adapted to manage with polynomials of several different orders. In practice, since at each step the actual order of terms associated to detuning is lower than the corresponding effective order, the algorithm is adapted by incorporating routines already used at previous steps. 

In practice, at step say $j$, we have an equation of the form
\be
K_j = {\cal H}_j + A_j + \delta B_{j-d} + \delta^2 B_{j-2d} + ... + L_{g_j} ({\cal H}_0),\ee
where $A_i, B_i$ are homogeneous polynomials of degree $i+2$ coming from previous steps. As usual, the algorithm is designed to identify in all terms with the exclusion of 
\be
L_{g_j}({\cal H}_0) \equiv -L_{{\cal H}_0} (g_j),\ee
monomials in the kernel of the linear operator $L_{{\cal H}_0}$. These monomials are used to construct $K_j$: the remaining terms are used to find $ g_j $ in the standard way. It is clear that both the normal form and the generating function are affected by the effective order of the detuning term.

In both cases (\ref{v1},\ref{v2}) investigated \cite{PBB}, with the detuning treated as a term of order 2, the next appearance of a related term is in $K_6$. Rather, if it is treated as a term of order 4, the next appearance of a related term is in $K_8$. Truncating at order 6 (polynomials of degree 8) is therefore sufficient to make a comparison with other predictions not sensitive to the detuning.

\subsection{Structure of the normal form}
\label{subsec2:4}

In principle, the recursion process to solve the system (\ref{EHK}) can be carried out to arbitrary order. In practice we have to truncate it at some finite order $N$. The ideal choice would be an {\it optimal} truncation order $N_{\rm opt}$, to evaluate which one can work with the formal integral (\ref{PHI}): minimizing its failure to commute with the Hamiltonian, one truncates the series at the order giving the best conservation \cite{giorgilli,CEG, ECG}. However, in general this is a costly procedure. 

On the other hand, a very conservative strategy can be that of truncating at the lowest order adequate to convey some non-trivial information on the system. In the resonant case, it can be shown \cite{tv} that the lowest order to be included in the normal form in order to {\it capture} the main effects of the $m_1/m_2$ resonance with double reflection symmetries is $N_{\rm min} = 2\times(m_1+m_2-1)$. Truncating at this level is enough to study the resonance and the main periodic orbits associated to it \cite{BBP1,BBP2}. Using `action-angle'--{\it like} variables $\vec{J}, \vec{\theta}$ defined through the transformation
\ba\label{AAV}
X &=& \sqrt{2 J_1} \cos \theta_1,\quad
P_X = \sqrt{2 J_1} \sin \theta_1,\\
Y &=& \sqrt{2 J_2} \cos \theta_2,\quad
P_Y = \sqrt{2 J_2} \sin \theta_2,\ea 
the typical structure of the doubly-symmetric resonant normal form truncated at $N_{\rm min}$ is \cite{c1,SV}
\be\label{GNF}
K=m_{1} J_1+m_{2} J_2+ \sum_{k=2}^{m_1+m_2} {\cal P}^{(k)}(J_1,J_2)+
a J_1^{m_{2}} J_2^{m_{1}} \cos [2(m_{2} \theta_{1}- m_{1} \theta_{2})], \ee
where ${\cal P}^{(k)}$ are homogeneous polynomials of degree $k$ whose coefficients may depend on $\delta$ and $a$ is a constant. In these variables the second integral is 
\be\label{cale}
{\cal E}=m_{1} J_1+m_{2} J_2\ee
and the angles appear only through the resonant combination
\be
\psi=m_{2} \theta_{1}- m_{1} \theta_{2}.\ee
Introducing its conjugate variable
\be\label{calr}
{\cal R}=m_{2} J_1-m_{1} J_2,\ee
the new Hamiltonian can be expressed in the {\it reduced} form $K({\cal R}, \psi; {\cal E},\delta)$ that is a family of 1-dof systems parametrized by ${\cal E}$ and $\delta$. 

\section{Applications}
\label{sec:3}

We now analyze a sample of problems that can be addressed and solved with the tools developed so far. The main periodic orbits and the families of quasi-periodic orbits parented by them give the structure of the phase space, at least in the regular regime. The study of existence and stability of normal modes and periodic orbits in general position admitted by (\ref{GNF}) or its higher-order generalization is therefore of outmost importance.

\subsection{Stability of normal modes}
\label{subsec3:1}

In systems of the form (\ref{Horig}) the orbits along the symmetry axes are simple periodic orbits. It can be readily verified that these orbits correspond to the two solutions in which either $J_1$ or $J_2$ vanish. If the axial orbit is stable it parents a family of `box' orbits. A case that is both representative of the state of affairs and useful in galactic applications is that of the stability of the  {\it x}-axis periodic orbit (the `major-axis orbit', if $q$ is in the range (\ref{rangeq})). Among possible bifurcations from it, the most prominent is usually that due to the 1:2 resonance between the frequency of oscillation along the orbit and that of a normal perturbation, producing the `banana' and `anti-banana' orbits \cite{mes}. The inclusion of detuning allows one to catch the passage of the system through the resonance due to the nonlinear coupling between the two degrees of freedom: the strength of the coupling depends on energy and we expect that the onset of the resonance is described by one (or more) curves on the $(\delta,E)$-plane. To investigate this problem in the potentials (\ref{v1},\ref{v2}), we construct the normal form with $m_{1}=1,m_{2}=2$ and study the nature of the critical points of the function $K^{(\mu)}=K+\mu {\cal H}_0$,
where $\mu$ has to be considered as a {\it Lagrange multiplier} to take into account that there is the constraint $ {\cal H}_0 = {\cal E} $. The condition for a change in the nature of the critical point corresponding to the normal mode is given by the solutions of the algebraic equation
\be
{\rm det}[{\rm d}^{2}K^{(\mu)}({\cal E},\delta)] |_{J_2=0}=0\ee
of degree $ N $ in ${\cal E}$: each transition of the kind extremum $\to$ saddle is equivalent to the onset of an instability and to the bifurcation of the banana (or of the anti-banana).  

However, in order to get a form usable in comparison with other results (for example coming from a numerical treatment) it is necessary to use a `physical' energy variable rather than the parameter ${\cal E}$. The conversion is possible if the physical energy $E$ appears explicitly \cite{BBP2}. According to the rescaling (\ref{newE}), we assume that $ m_{2} q E $ is the constant `energy' value assumed by the truncated Hamiltonian $K$. In the present instance $m_{2}=2$ so that, on the {\it x}-axis orbit, the new Hamiltonian is a series of the form
\be\label{KIa}
K = 2 q {\cal E} + c q {\cal E} ^{2}+ ... = 2qE.\ee
This series can be inverted to give
\be
{\cal E}= E - \frac{c}2 E^2 + ...\ee
and this can be used in the treatment of stability to replace ${\cal E}$ with $E$. 
Recalling that, in this case, (\ref{DET}) gives $q=1/2+\delta$, every solution can be expanded as
\be\label{sqe}
E_{\rm crit}(\delta) = \sum_{k=1}^{N/2} c_{k}{\delta}^{k}
\ee
and in this form they can be used for quantitative predictions. 

 \begin{table}
 \centering
    \caption{Coefficients in the expansion (\ref{sqe}) with $N=14$ for the logarithmic potential (banana, 2nd column and anti-banana, 3rd column) and the conical potential (banana, 4th column and anti-banana, 5th column).}
\label{tl12}
    \begin{tabular}{||c||c|c||c|c||}
      \hline
      $$ & \multicolumn{2}{c||} {{\rm Potential} $V_{L}$} & \multicolumn{2}{c||} {{\rm Potential} $V_{C}$} \\ \hline
 $k$ & ${\rm Banana}$ &      ${\rm Anti-banana}$ & $ {\rm Banana} $  &   $ {\rm Anti-banana} $ \\
									   \hline 
 $1$ & $ 8 $  &   $ 8 $ & $ 16 $  &   $ 16 $ \\ 
& &  & & \\
 $2$ & $ -{{20}\over{3}} $  &   $ {{28}\over{3}} $ & $ {{248}\over{3}} $  &   $ {{536}\over{3}} $\\ 
 & &  & & \\
 $3$ & $ \frac{268}{9} $  &   $ \frac{460}{9} $ & $ \frac{3608}{9} $  &   $ \frac{18584}{9} $\\
 & &  & & \\
 $4$ & $ -\frac{1724}{27} $  &   $ \frac{3928}{27} $ & $ \frac{43328}{27} $  &   $ \frac{657848}{27} $ \\
 & &  & & \\
 $5$ & $ \frac{79184}{405} $  &   $ \frac{267404}{405} $ 
        & $ \frac{525704}{81} $  &   $ \frac{23668304}{81} $\\
 & & & & \\
 $6$ & $ -\frac{567178}{1215} $  &   $ -\frac{510200857}{405} $ 
        & $  \frac{28118794}{1215} $  &   $ \frac{4304374384}{1215} $ \\
 & & & & \\
 $7$ & $ -{{30991946}\over{25515}} $  &   $ {{615376795556}\over{8505}} $ 
        & $  {{309430864}\over{3645}} $  &   $ {{31575390356}\over{729}} $ \\
 \hline
\end{tabular}
\end{table}

In Table \ref{tl12}. we list the coefficients of the series (\ref{sqe}) giving these bifurcations for the logarithmic potential (\ref{v1}) and the `conical' potential  (\ref{v2}). They have been obtained \cite{PBB} with a normal form truncated at order $N=14$ and with the detuning treated as a term of order 2. There is a complete agreement with the analytical approach based on the Poincar\`e-Lindstedt method \cite{scu} and, as discussed below, there is a striking agreement with the numerical approach based on the Floquet method. The agreement of all fractional coefficients is complete up to $N/2 = 7$. On the other hand, if the detuning is treated as a term of order $d = 4$ or greater, we get a disagreement in the coefficients starting from $c_{3}$. This result confirms the analysis made above on the `propagation' of the detuned terms in the normal form and show that the choice $d=2$ is the optimal one. 

 \begin{table}
 \centering
   \caption{Subsequent truncations of expansion (\ref{sqe}) with $N=14$ for the logarithmic potential (banana). $E_{B}$ is the value obtained by means of the Floquet method.}
\label{pl12}
    \begin{tabular}{||c||c|c|c|c||}
      \hline
      $$ & \multicolumn{4}{c||} {$\delta$} \\ \hline
 $n$ & $0.1$ &      $0.2$ & $ 0.3 $  &   $ 0.4 $ \\
									   \hline 
 $1$ & $ 0.800000 $  &   $ 1.60000 $ & $ 2.40000 $  &   $ 3.20000 $ \\
& & 	& &								   \\
 $2$ & $ 0.733333 $  &   $ 1.33333 $ & $ 1.80000 $  &   $ 2.13333 $ \\
& & 	& &								   \\
 $3$ & $ 0.763111 $  &   $ 1.57156 $ & $ 2.60400 $  &   $ 4.03911 $ \\
 & & 	& &								   \\
 $4$ & $ 0.756726 $  &   $ 1.46939 $ & $ 2.08680 $  &   $ --- $ \\
 & & 	& &								   \\
 $5$ & $ 0.758681 $  &   $ 1.53196 $ & $ 2.56190 $  &   $ --- $ \\
 & & 	& &								   \\
 $6$ & $ 0.758214 $  &   $ 1.50208 $ & $ 2.22160 $  &   $ --- $ \\
 & & 	& &								   \\
 $7$ & $ 0.758336 $  &   $ 1.51763 $ & $ 2.48724 $  &   $ --- $ \\
        \hline
        $E_{B}$ & $ 0.758 $  &   $ 1.513 $ & $ 2.401 $  &   $ 3.646 $ \\
 \hline
\end{tabular}
\end{table}

What is remarkable in the quality of the prediction with regard to `experimental' numerical data is that numerical computations are performed with the {\it exact} logarithmic (or conical) potentials (\ref{v1},\ref{v2}), whereas the analytical predictions are, in any case, based on the series expansions of these potentials that appear in (\ref{detH}) with limited convergence radii. The reliability of these predictions in a range wider than foreseen can be explained if we interpret the series of the form (\ref{sqe}) as {\it asymptotic series} and evaluate their truncations by computing the successive partial sums
\be
E_{n}(q) = \sum_{k=1}^{n} c_{k}{\delta}^{k}, \quad n=1,...,N/2.
\ee
Minimizing the difference between the `exact' value and its approximations provides an estimate of the optimal truncation. As an example, in Table \ref{pl12}. we report these partial sums for the banana bifurcation in the logarithmic potential (\ref{v1}), with $0.1<\delta<0.4 \; (0.6<q<0.9)$ and compare them with the values obtained by means of the Floquet method \cite{mes,BBP2} given in the last row. The numerical values of the partial sums are given with 6 digits just to show more clearly the asymptotic behaviour: we can see that, up to $\delta=0.3$, the predictions are apparently still (slowly) converging at $n=7$. Only at the rather extreme value $\delta=0.4$ we get an `optimal' truncation order $n_{\rm opt} = 3$, with a $10 \%$ error on the exact value of the critical energy. We may wonder if there is a way to speed up the convergence rate: this can be done with a resummation method like the continued fraction \cite{PBB}. It can be shown that, for all values of $\delta$ up to $0.3$, $n=6$ {\it is enough to reach a precision comparable to the numerical error}. For $\delta=0.4$ we get an optimal truncation order $n_{\rm opt} = 5$, with a $3 \%$ error on the exact value of the critical energy.

\subsection{Periodic orbits in general position and boxlets}
\label{subsec3:2}

In addition to the normal modes, each resonant normal form of the type (\ref{GNF}) admits a double family of resonant periodic orbits in general position usually called {\it boxlets} in galactic dynamics \cite{mes}. They can be easily identified using the fact that the two `angles' have a fixed phase relation given either by $\psi=0$ or $\psi=\pm \pi$. In addition to the already mentioned 1:2 resonant banana, we have the 1:1 `loop', the 2:3 `fish' and so forth \cite{mes}. Each of them, if stable, is surrounded by a family of quasi-periodic orbits usually inheriting the same nickname. In a given potential in the class of (\ref{Horig}), several boxlets can be present at the same time, whereas each resonant normal form is able to correctly render only one type: even with this limitation, a knowledge of the corresponding family is very useful. In particular, we can analytically compute the phase-space fraction occupied by the given family, an important information in the process of constructing self-consistent models. We illustrate the idea in the case of the loop family: the principle is the same for higher resonating boxlets but the computations much more involved.

Usually the loop bifurcates from the minor-axis periodic orbit at energy lower than that of the banana bifurcation fron the major axis \cite {BBP1}: there is a regime in which the loop family and the boxes around the stable major-axis orbit coexist. To identify loops, we impose the condition that the Hamiltonian flow generated by the 1:1 version of the reduced normal form $K({\cal R}, \psi; {\cal E},\delta)$ has a fixed point in ${\cal R} = {\cal R}_{L},\psi= \pi$. Using relations (\ref{cale}) and (\ref{calr}) and the value of the detuning $\delta = q-1$, this solution fixes the actions on the closed loop: $J_{1L}({\cal E},q)$ and $J_{2L} = {\cal E}-J_{1L}({\cal E},q)$. On the periodic orbit, it is possible to find a relation between ${\cal E}$ and the true energy $E$ in a form analogous to the expansion (\ref{KIa}). We can then express the actions as a series in $E$ and, exploiting their geometric meaning, produce an estimate of the fraction of phase space occupied by the loops {\it and} the boxes. Truncating the series at first order, in the logarithmic case the results are \cite {BBP2} 
\be
f_{Loop} = \frac{J_{1L}(E,q)}{{\cal E}(E)} = 
                 \frac{2(-3 + 3 q - 5 q^2 + 5 q^3) + E (9 - 9 q + 11 q^2 - 3 q^3)}
                       {(3 - 2 q + 3 q^2)(-2(1-q)^{2}+ E (3 - 2 q + 3 q^2))} \ee
and $f_{Box} = 1-f_{Loop}$. These predictions and the corresponding one for the banana family \cite {BBP2} agree very well with numerical estimates \cite{mes} up to energy values much greater than that corresponding to the harmonic core of the potential.

\subsection{Singular limits}
\label{subsec3:3}

The potentials considered up to now are assumed to be analytic in the origin. However, we know that realistic models should include a singularity related to a density `cusp' and/or a central point mass \cite{ss}. In the examples of the form (\ref{v1},\ref{v2}) is implicitly assumed the use of adimensional coordinates by introducing a `core radius' $R_{c}$ which can be put equal to $1$ without less of generality. In the limit $R_{c}\to0$, those examples reduce to members of the familiy of {\it singular scale-free} potentials \cite{tt}
 \be
 V_{\alpha} (s) = A \, s^{\alpha}, \quad -1/2 \le \alpha \le 1, \quad A \alpha > 0.\ee
 The singular conical potential is given by $\alpha=1/2$ and $A=1$ while the singular logarithmic potential corresponds to the limit $\alpha\to0$ with $A=1/2$. 

It is tempting to try to extract information concerning the scale-free singular limit from our analytical setting based on series expansions. Formally, this operation should be hindered by the lack of a series representation of the singular potential. However, we may nonetheless `force' our approximate integrals of motion to play their role in the singular limit too and try our chance by constructing a Poincar\'e surface of section by using the approximate integral $I (x,y,p_x,p_y;q)$ given by (\ref{eqn:In}). In view of the scale invariance, we fix the energy level $E_{0}$ and construct, e.g., a $y$-$p_y$ surface of section by means of the intersection of the function $I (x,y,p_y;E_{0},q)$ with the $x=0$ hyperplane. The level curves of the function
\be\label{FF}
F (y,p_y) = I (0,y,p_y;E_{0},q)\ee
give the invariant curves on the section. In the singular logarithmic case with $E_{0}=0$ and $q=0.7$ 
we get, quite surprisingly, acceptable results \cite{BBP2}. In the section constructed by using the approximate integral $I^{(1:1)}$ related to the 1:1 resonance and obtained by truncating the series (\ref{eqn:In}) at order 6, it gives the family of loops around the stable periodic orbit at $y \simeq 0.56 $ in good agreement with numerical data \cite{mes}. A similar result is given with the section obtained by using the approximate integral $I^{(1:2)}$ again truncated at order 6: it gives the family around the stable banana at $y \simeq 0.16$ and boxes around it. 

\section{Comments and outlook}
\label{sec:4}

As any analytical approach, this
method has the virtue of embodying in (more or less) compact formulas simple rules to compute specific quantities, giving a general overview of the behavior of the system. In the case in which a non-integrable system has a regular behavior in large part of its phase space, a very conservative strategy, like that of truncating at a low order including the resonance, provides sufficient qualitative and quantitative agreement with other more accurate but less general approaches. In our view, the most relevant limitation, common to all perturbation methods, is due to the intrinsic structure of a {\it single-resonance} normal form. However, we remark that the regular dynamics of a non-integrable system can be imagined as a superposition of very weakly interacting resonances. If we are not interested in thin stochastic layers, each portion of phase space associated with a given resonance has a fairly good alias in the corresponding normal form. 

There are several lines of developement of this line of research; we mention a few of them: to extend the asymptotic analysis of perturbation series representing the building blocks of phase space (actions, frequencies, etc.); to devise suitable coordinate transformations to enable the investigation of cuspy potentials and/or central `black holes'; to apply the normalization algorithm to three degrees of freedom systems with and without rotation.

\section*{Acknowledgment}

I wish to thank George Contopoulos, Christos Efthymiopoulos, Giuseppe Gaeta, Luigi Galgani, Antonio Giorgilli and Ferdinand Verhulst for very useful discussions.

\printindex
\end{document}